\documentclass[aps,pre,superscriptaddress,floatfix]{revtex4-2}
\usepackage{amsmath}
\usepackage{graphicx}
\usepackage{subfigure}
\usepackage{dcolumn}
\usepackage{epstopdf}
\usepackage{natbib}
\usepackage{stfloats}
\usepackage{epsfig,amssymb,subfigure,bm,dsfont}
\usepackage{lipsum} 
\usepackage{appendix}
\usepackage{float}
\usepackage{tikz}
\usetikzlibrary{arrows,shapes,chains}

\usepackage[colorlinks, bookmarks=true,breaklinks=true,linkcolor=red, citecolor=blue, linktocpage=true, urlcolor=blue]{hyperref}
\renewcommand\appendix{\par
	\setcounter{section}{0}
	\setcounter{subsection}{0}
	\gdef\thesubsection{\Roman{subsection}}}
\begin{document}

	\preprint{APS/123-QED}
	\title{Perturbing Chaos with Cycle Expansions}
	\author{Huanyu Cao}
	\affiliation{School of science, Beijing University of Posts and Telecommunications, Beijing 100876, China}
	\author{Yueheng Lan}
	\email{lanyh@bupt.edu.cn}
	\affiliation{School of science, Beijing University of Posts and Telecommunications, Beijing 100876, China}
	\affiliation{State Key Lab of Information Photonics and Optical Communications, Beijing University of Posts and Telecommunications, Beijing 100876, China}
	
	\begin{abstract}
		\noindent\textbf{Abstract.} 
		Due to existence of periodic windows, chaotic systems undergo numerous bifurcations as system parameters vary, rendering it hard to employ an analytic continuation, which constitutes a major obstacle for its effective analysis or computation. In this manuscript, however, based on cycle expansions we found that spectral functions and thus dynamical averages are analytic, if symbolic dynamics is preserved so that a perturbative approach is indeed possible. Even if it changes, a subset of unstable periodic orbits (UPOs) can be selected to preserve the analyticity of the spectral functions. Therefore, with the help of cycle expansions, perturbation theory can be extended to chaotic regime, which opens a new avenue for the analysis and computation in chaotic systems.
	\end{abstract}
	\maketitle

	\section{\label{sec:Introduction}Introduction}
		
		Turbulent systems often exhibit characteristic recurrent patterns, which are routinely observed in both numerical simulations and wet experiments and are termed coherent structures. These recurrent patterns are compact invariant sets with relatively simple topology in phase space~\cite{kline1967structure,liepmann2001elements,cvitanovic2010geometry} and dominate dynamics of fluid systems~\cite{cvitanovic1991periodic2,cvitanovic2010geometry}. Intuitively, at finite resolution, the spatiotemporal evolution can be regarded as a walk through the labyrinth of finitely many unstable periodic orbits (UPOs, also called cycles). Such a view enables a hierarchical description of the fluid motion as demonstrated in cycle expansions~\cite{cvitanovic2005chaos}. These cycles are locally well organized and accessible through analytical approximation or numerical computation, which provides the desired skeleton of the irregular dynamics as mentioned above. 
		
		The importance and properties of UPOs have been emphasized ever since Poncar{\'e}'s work on dynamical systems, for they carry both topological and dynamical information~\cite{cvitanovic2005chaos}. From the perspective of physical intuition, a trajectory in a chaotic system always evolves adjacent to a UPO for some time, and then sticks to another UPO for some time, and so on~\cite{lan2010cycle}. The UPOs act as the ``skeleton'' of the system, which could be organized in a hierarchical manner~\cite{cvitanovic1991periodic}. The POT supplies a formalism of relating dynamical averages to the spectra of appropriate evolution operators with the natural measure as the eigenstate corresponding to the leading eigenvalue. The trace and spectral determinant of the evolution operator are defined which can be evaluated with UPOs and dynamical averages are expressible in terms of their eigenvalues~\cite{cvitanovic2005chaos}. As a result, an average over the natural measure is expressed in terms of the corresponding quantity evaluated on UPOs. Cycle expansions is an efficient way to reveal the shadowing property embedded in system dynamics which efficiently procures the spectrum of evolution operators with cycles. For nice hyperbolic systems, the spectral determinant and the dynamical zeta function turns out to be analytic in a neighborhood of $z=0$ and the cycle expansion technique re-expresses them in terms of a convergent sum over UPOs ordered in a hierarchical way with corrections from long cycles declining rapidly.
		
		All the previous discussions are concentrated on the unperturbed systems whose state evolution is governed by given maps or differential equations. Nevertheless, in realistic experiments, the deterministic behaviour or fixed dynamics is only an idealization since noise and perturbations are inevitable. Under perturbation, chaotic systems may retain chaoticity or enter different regimes of dynamics~\cite{kaplan1979preturbulence}. During the process, very often bifurcations are observed and it is possible that small perturbation may induce qualitative changes in global dynamics, which is best exemplified in chaos control~\cite{braiman1991taming,vohra1995suppressed,gonzalez1998resonance} or transient chaos maintenance~\cite{in1995experimental,schwartz1996sustaining}. However, the quantitative analysis of chaotic systems subject to perturbations are hard to carry out due to these "unpredictable bifurcations" which exist densely in the parameter space. 
		
		Essentially, the basis for performing a quantitative analysis with perturbation scheme is analyticity or continuity of the sought solutions. A specific cycle changes smoothly upon parameter variation until disappears at some bifurcation point. However, infinitely many unstable cycles exist in a chaotic system, part of which always changes qualitatively when system parameters shift. Therefore, the average of an observable in general is not an analytic function of parameters since all the cycles have to be included in its computation based on cycle expansions. Nevertheless, for a computation with finite accuracy, only finitely many cycles are used. If their creation or annihilation can be tracked during the whole process, analyticity may be recovered on the relevant subset of cycles.  In this paper, we focus on a perturbative computation of observable averages with these cycles based on cycle expansions. In one or two dimensions, symbolic dynamics could be used to monitor the existence of cycles. If no bifurcation occurs during parameter shift, the observable average is an analytic function of the parameter and a simple Taylor expansion may be employed. If bifurcations do occur, only a subset of cycles may be used to compute expansion coefficients. Several examples are utilized to demonstrate the validity of the current scheme. It turns out that this combination of qualitative analysis based on symbolic dynamics and quantitative computation with cycle expansion is indeed able to provide a new tool to cross bifurcations and recovers a perturbative investigation of chaotic systems.

		The paper is organized as follows: in Sect.~\ref{sec:POT and SD}, we briefly review the related contents of POT and symbolic dynamics. In Sect.~\ref{sec:Perturbation Scheme}, our perturbation scheme based on cycle expansions is introduced in detail for predicting observable averages. In addition, some necessary pruning rules and algorithms are discussed. In Sect.~\ref{sec:Examples}, several 1- and 2-dimensional models are applied to demonstrate the effectiveness of our scheme, and then we conclude with a summary and a vision of future developments in Sect.~\ref{sec:Summary}. Some details are included in the appendix.

	\section{\label{sec:POT and SD}Periodic Orbit Theory and Symbolic Dynamics}
	
		\subsection{\label{subsec:POT}Periodic Orbit Theory}
		
		In chaotic systems, it is difficult to track long-term evolution of individual trajectories due to sensitivity to initial conditions so we focus on statistical properties of chaotic systems instead, i.e. the averages of certain observables. The phase space of a chaotic system is densely covered with UPOs, which could be conveniently used to compute these averages. Here, we do not pursue mathematical rigor but rather emphasize physical intuitions and practical applications. From a statistical physics perspective, POT actually provides a method to accurately extract the information we are interested in with a series of UPOs and is powerful for reliable and accurate analysis in hyperbolic chaotic systems. 
		
		Although the method is applicable to both continuous and discrete time evolution, here we only discuss discrete dynamics for brevity without loss of generality. Very often, the time average of an observable $a(x)$ can be evaluated~\cite{cvitanovic2005chaos} along a trajectory from an arbitrary typical initial point $x_0$ in phase space $\mathcal{M}$
			\begin{eqnarray}{\label{Formula:timeaverage}}
				\bar{a}_{x_0}=\lim\limits_{n\to\infty}\frac{A^n}{n}=\lim\limits_{n\to\infty}\frac{1}{n}\sum_{k=0}^{n-1}a(f^{k}(x_0))
				\,,
			\end{eqnarray}
		where $x_{n+1}=f(x_n)$ describes the dynamics of the given system and $A^n(x_0)=\sum_{k=0}^{n-1}a(f^{k}(x_0))$ is defined as the integrated observable~\cite{cvitanovic2005chaos}. In practical computation, however, a time average can only be approximated with a finite number of iterations. An alternative scheme is to compute the weighted spatial average~\cite{cvitanovic2005chaos}. If a normalized measure $\omega(x)$ exists in the phase space $\mathcal{M}$, the weighted spatial average could be defined as
			\begin{eqnarray}{\label{Formula:phaseaverage}}
				\langle a\rangle_{\omega}=\int_{\mathcal{M}}a(x)\omega(x)dx
				\,.
			\end{eqnarray}
		As time tends to infinity, any typical initial measure evolves to an asymptotic measure $\rho(x)$, being named natural measure~\cite{cvitanovic2005chaos,cvitanovic1988invariant}. If the dynamics is ergodic, a natural measure $\rho(x)$ exists for which the two averages are equal, {\em i.e.} $\langle a\rangle_{\rho}=\bar{a}_{x_0}$ for almost all initial point $x_0$. Generally, it is difficult to obtain an explicit expression for the natural measure defined on a fractal set characteristic of a strange attractor in chaotic dynamics. 
		Fortunately, POT provides new insight into capturing the generally elusive natural measure. In brief, dynamical features can be extracted through a well-designed evolution operator $\mathcal{L}^n$~\cite{cvitanovic2005chaos}, which is defined as
			\begin{eqnarray}{\label{Formula:evolution operator}}
				\mathcal{L}^n\circ\omega(y)=\int_{\mathcal{M}}dx\delta(y-f^n(x))e^{\beta A^n(x)}\omega(x)
				\,,
			\end{eqnarray}
		where $n=1,2,\cdots$ for discrete mappings. The kernel function $\mathcal{L}^n(y,x)=\delta(y-f^n(x))e^{\beta A^n}$ depends on the integrated quantity $A^n$ and an auxiliary variable $\beta$. That is, the evolution operator is able to describe the evolution of the measure $\omega(x)$ and to record the integrated observable along an orbit. If we set $\beta=0$, $\mathcal{L}$ is the famous Perron-Fr{\" o}benius operator~\cite{cvitanovic2005chaos}. 
		
		Denoting the spectrum of $\mathcal{L}$ by $\{s_m\}_{m\in\mathbb{N}}$ with $\mathrm{Re}(s_m)>\mathrm{Re}(s_{m+1})$. From the perspective of spectral considerations, high powers of the linear operator $\mathcal{L}$ are dominated by the leading eigenvalue $s_0$, specifically
			\begin{eqnarray}{\label{Formula:trace approximation}}
				\mathcal{L}^n\circ I(x) =\sum_{m}b_m\phi_m(x)e^{ns_m}\to b_0\phi_0(x)e^{ns_0},\,n\to\infty
				\,,
			\end{eqnarray} 
		where $I(x)\equiv 1/\langle1\rangle_{I}$ is the identity function and expressed as an expansion of the eigenfunctions $\phi_m(x)$ of $\mathcal{L}$, {\em i.e.}, $I(x)=\sum_m b_m \phi_m(x)$. Thus, in terms of the evolution operator, we have 
			\begin{eqnarray}{\label{Formula:<ebetaAn>}}
				\langle e^{\beta A^n}\rangle_{I}=\int_{\mathcal{M}}dx[\mathcal{L}^n\circ I](x)\to b_0e^{ns_0},\,n\to\infty
				\,,
			\end{eqnarray} 
		where $s_0$ is a function of $\beta$ and thus we have 
			\begin{equation}\label{Formula:s0}
				s_0(\beta)=\lim\limits_{n\to\infty}\frac{1}{n}\ln(\langle e^{\beta A^n}\rangle)_{I} 
				\,.
			\end{equation}
		If the system is ergodic, the average
			\begin{equation}{\label{Formula:<a>=ps/pbeta}}
				\langle a \rangle=\lim\limits_{n\to\infty}\frac{1}{n}\langle  A^n\rangle_{I}=\frac{d s_0(\beta)}{d \beta}|_{\beta=0}
			\end{equation}
		is directly related to the leading eigenvalue $s_0(\beta)$. So, all we need to do is extract the spectrum of $\mathcal{L}$, especially the leading eigenvalue $s_0$. 
		
		The spectrum of the linear operator $\mathcal{L}$ is determined by solving the resolvent equation $\mathrm{det}(\mathbf{1}-z\mathcal{L})=0$. Borrowing the identity between the determinant and trace of an arbitrary square matrix $M$: $\mathrm{ln}\,\mathrm{det}\,M=\mathrm{tr}\,\mathrm{ln}\,M$, we have the spectral determinant~\cite{cvitanovic2005chaos,artuso1990recycling1,artuso1990recycling2}
		\begin{align}{\label{Formula:spectral determinant}}
		\mathrm{det}(\mathbf{1}-z\mathcal{L})&=\mathrm{exp}(\mathrm{tr}\,\mathrm{ln}(\mathbf{1}-z\mathcal{L}))=\mathrm{exp}\left(-\sum_{n=1}^{\infty}\frac{z^n}{n}\mathrm{tr}\mathcal{L}^n\right)\nonumber\\
		&=\mathrm{exp}(-\sum_{p}\sum_{r=1}^{\infty}\frac{1}{r}\frac{z^{n_pr}e^{r\beta A_p}}{|\mathrm{det}(\mathbf{1}-\mathit{M}_p^r)|})
		\,,
		\end{align}
		where $p$ denotes prime cycles which are not repeats of shorter ones and $n_p$ is the length of the cycle $p$. $A_p$ and $M_p$ are the integrated physical quantity and the Jacobian matrix along the prime cycle $p$. The trace $\mathrm{tr}\,\mathcal{L}^n$ in the above equation has been computed with the trace formula~\cite{cvitanovic2005chaos,artuso1990recycling2}
		\begin{align}{\label{Formula:trace formula}}
			\mathrm{tr}\mathcal{L}^n=&\int_{\mathcal{M}}dx\mathcal{L}^n(x,x)=\int_{\mathcal{M}}dx\delta(x-f^n(x))e^{\beta A^n}\nonumber\\
			=&\sum_{f^n(x_i)=x_i}\frac{e^{\beta A^n(x_i)}}{|\mathrm{det}(\mathbf{1}-M_n(x_i))|},\forall n\in\mathbb{Z}^+
			\,,
		\end{align}
		where $x_i$ is a periodic point of period $n$ and $M_n(x_i)$ is the Jacobian matrix of $f^n(x)$ evaluated at $x_i$. Based on the hyperbolicity assumption~\cite{cvitanovic2005chaos} that the stabilities of all cycles included in Eq.(\ref{Formula:spectral determinant}) are exponentially bounded away from unity, we make the approximation $1/|\mathrm{det}(\mathbf{1}-M_p^r)|\approx1/|\Lambda_p|^r$, where $\Lambda_p=\prod_{e}\Lambda_{p,e}$ is the product of expanding eigenvalues of the matrix $M_p$. With $r\to\infty$, the spectral determinant Eq.(\ref{Formula:spectral determinant}) becomes the dynamical zeta function~\cite{cvitanovic2005chaos,cvitanovic1999spectrum}
		\begin{eqnarray}{\label{Formula:dynamical zeta function}}
			\frac{1}{\zeta}=\prod_{p}(1-t_p),
		\end{eqnarray}
		where $t_p=\frac{z^{n_p}e^{\beta A_p}}{|\Lambda_p|}$ denotes the weight of prime cycle $p$. It can be proved that the dynamical zeta function is the $0$th-order approximation of the spectral determinant and they have identical leading eigenvalue but different analytic properties~\cite{cvitanovic2005chaos,artuso1990recycling1}. 
		
		For a chaotic system satisfying the hyperbolicity assumption, a long cycle is often well  approximated with several shorter ones, which is indicated by the shadowing lemma in nonlinear dynamics~\cite{cvitanovic2005chaos}. Based on this property, cycle expansion is designed to efficiently deal with the spectral functions Eq.(\ref{Formula:spectral determinant}) or Eq.(\ref{Formula:dynamical zeta function}), with short periodic orbits capturing the major part of the natural measure and longer cycles delivering systematic curvature corrections. For maps with binary symbolic dynamics~\cite{artuso1990recycling2}, Eq.(\ref{Formula:dynamical zeta function}) is expanded as
		\begin{eqnarray}{\label{Formula:cycle expansion}}
			\begin{split}
			\frac{1}{\zeta}=&1-\sum_{f}t_f-\sum_{p}c_p=1-t_0-t_1-[(t_{01}-t_0t_1)]\\
			-&[(t_{001}-t_{01}t_0)+(t_{011}-t_{01}t_1)]-...,
			\end{split}
		\end{eqnarray}
		where the fundamental terms $t_f$ include all unbalanced, not shadowed prime cycles and the rest terms $c_p$, called curvature corrections, consist of longer prime cycles and pseudo-cycles that shadow them. Cycle expansions are dominated by fundamental terms, with long orbits contributions cancelled by short ones, so that curvature corrections decay exponentially or even super-exponentially if uniform hyperbolicity is assumed~\cite{artuso1990recycling1}. The cancellation between prime cycles and pseudo-cycles reflects the smoothness of the underlying dynamics~\cite{lan2010cycle}. 
		
		Very often in practical computation, a good truncation to the spectral functions is a crucial operation to restrict the computation within finitely many unstable cycles in a chaotic system. The usually adopted truncation with cycle length corresponds to a geometric envelope approximation of the original map~\cite{cvitanovic1991periodic2}. Compared with the reserved term, the magnitude of the discarded terms in the formula decreases exponentially with the topological length and higher order truncations lead to a more accurate evaluation. 

		However, most physical systems are not uniformly hyperbolic so that the cancellation is poor. One non-hyperbolicity case is marked with the strong contraction at specific locations of an attractor such as critical points in 1-d maps or homoclinic tangencies in the Hénon map~\cite{artuso1990recycling1,artuso1990recycling2}. As a consequence, there are singularities in the natural measure which undermine the shadowing and slow down the convergence of cycle expansions. Several accelerating schemes have been proposed, among which stability ordering is a good choice~\cite{dettmann1997stability}. It retains all the cycles or pseudo-cycles that have stability eigenvalues smaller than a threshold in the cycle expansion. The method is based on analyticity of the spectral functions~\cite{artuso1990recycling2,main1999semiclassical}, which identifies and removes the poles that are near the origin and thus expands the radius of convergence. With appropriate coordinate transformations, dynamical conjugacy can be used to remove the singularities in the natural measure and accelerate the convergence~\cite{gao2012accelerating}. In intermittent systems, the dynamics could alternate between regular and chaotic motion which results in non-hyperbolicity. The spectrum of the evolution operator is no longer discrete and the dynamical zeta function exhibit branch cut~\cite{artuso2003cycle}. Geometrically, the UPOs which have a stability eigenvalue close to 1 possess an unusually large weight and cannot be efficiently shadowed by shorter cycles~\cite{lan2010cycle}. A dynamics-splitting algorithm has been proposed to take advantage of the partial integrability of intermittent systems which analytically estimates the natural measure near the singularities but employ cycle expansions to treat the rest~\cite{cao2022wielding}. In the situations of interest in this paper, upon parameter change some UPOs may disappear and lead to bad-shadowing. Or, there is a quenched disorder in the dynamics and we need to do cycle expansion for many different parameter values. It will be shown below that the analyticity of the cycles could be used to carry out perturbation in the spectral functions. 
		\subsection{\label{subsec:SD}Symbolic Dynamics}
			
\begin{figure}[!htbp]	
			\centering
			\includegraphics[scale=0.5]{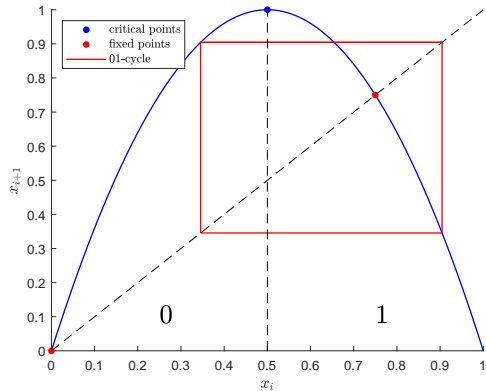}
			\caption{ The logistic map $f(x)$ (blue line) is defined on the unit interval $[0,1]$ and the critical point (blue dot) divides the phase space into two halves denoted by alphabet $\{0,1\}$ separately. The fixed points $x=0$ and $x=3/4$ (red dots) are actually period-1 orbits and represented by $\bar{0}$ and $\bar{1}$ respectively. The period-2 orbit (red line) has the sequence $0101...$ and is denoted by $\bar{01}$.}
			\label{graphic:logistic map}
		\end{figure}

		Symbolic dynamics~\cite{hao1998applied} is a very effective theory to divide and encode the whole phase space when searching for orbits or exploring the topological structure of dynamics. We introduce basic notions about it with the logistic map $x \mapsto f(x)=4x(1-x), x\in[0,1]$. We partition the phase space with the critical point $x_c=1/2$, and label the two non-overlapping intervals $[0,1/2)$ and $[1/2,1]$ with ``$0$'' and ``$1$'' respectively so that a trajectory is uniquely associated with a binary symbol sequence $x_0x_1x_2x_3..., x_i\in\{0,1\}$, called itinerary, according to the intervals which the trajectory consecutively visits. A good partition ensures that two different unstable trajectories have distinct itineraries. A family of orbits can be denoted as $x_0x_1x_2...x_{k-2}x_{k-1}$, which visit same intervals within $k$ iterations. A period-$m$ prime cycle is denoted as $\overline{x_0x_1...x_{m-1}}$ which is not repeats of shorter ones. For example, the period-$2$ cycle in Fig.~\ref{graphic:logistic map} is described by the infinite sequence $010101...$, which may be denoted as $\overline{01}$ and has a topological length of $2$. 
		Combining geometric thinking, it is feasible to establish criteria to identify inaccessible itineraries and thus detect all short prime cycles in a given system. In other words, we can rely on symbolic dynamics to sort the spatial orders of the prime cycles and search for admissible UPOs. In 1-dimensional cases, the kneading theory~\cite{cvitanovic2005chaos} (detailed in App.~\ref{appendix:SD}) provides a precise and definitive criterion of admissibility which eliminates all itineraries and UPOs that cannot occur for a given map. And in 2-dimensional cases, the kneading theory can be generalised to the so-called pruning front conjecture~\cite{cvitanovic1988topological} (detailed in App.~\ref{appendix:SD}), which offers a complete description of the symbolic dynamics of orientation reversing once-folding maps in the same sense as that the kneading sequence gives in a 1-dimensional unimodal map. In some cases, we may still find all the short admissible UPOs even without an elaborate pruning rule. Based on a good partition of the phase space and the associated mapping relation, admissible UPOs are found directly with cycle-detecting algorithms, although many symbol sequences do not match any admissible orbits.
	\section{\label{sec:Perturbation Scheme}Perturbation Scheme}
		
		\subsection{\label{subsec:Perturbed Model}Perturbed Model}
		As introduced in \cite{lan2010cycle}, in locally well organized flows, coherent structures interact weakly with each other except at some discrete space–time points where they are annihilated or created. Similar cellular subsystems may be simplified as a series of low-dimensional models with parameters selected from a given distribution~\cite{deane1991low,dankowicz1996local} and are ready for thorough analytical or numerical investigation. On this occasion, it is essential to study a series of chaotic systems with similar structure which may be treated in batch with a specifically designed perturbation theory.
		
 		Without loss of generality, we consider a model $f(x)$ defined in $\mathcal{M}$ under a given perturbation being expressed as 
 		\begin{eqnarray}{\label{Formula:the perturbed map}}
 			f_{\epsilon}(x)=f(x)+\epsilon g(x),x\in\mathcal{M},|\epsilon|\ll\mathcal{O}(1)\,,
 		\end{eqnarray}
 		where $g(x)$ defines the form of the perturbation and $\epsilon$ indicates its strength. Obviously, as long as $f_{\epsilon}(x)$ retains hyperbolicity, a fast convergence of cycle expansion results no matter what form $g(x)$ is. Thus we have the perturbed dynamical zeta function
 		\begin{eqnarray}{\label{Formula:perturbed dynamical zeta function}}
 		\frac{1}{\zeta}_{\epsilon}=\prod_{p}(1-t_{p,\epsilon}), t_{p,\epsilon}=\frac{z^{n_p}e^{\beta A_{p,\epsilon}}}{|\Lambda_{p,\epsilon}|}.
 		\end{eqnarray}	
 		
 		For convenience, we denote $\frac{1}{\zeta}_{\epsilon}$ as $F_{\epsilon}(s_{0,\epsilon}(\beta),\beta)$ where $s_{0,\epsilon}$ is the leading eigenvalue of perturbed evolution operator $\mathcal{L}_{\epsilon}$ and $F_0\equiv F(s_{0,0}(\beta),\beta)$ is the unperturbed dynamical zeta function. According to Eq.(\ref{Formula:<a>=ps/pbeta}), observable averages can be computed through the derivatives of $F_{\epsilon}(s_{0,\epsilon}(\beta),\beta)$~\cite{cvitanovic2005chaos}
 		\begin{eqnarray}{\label{Formula:average computing}}
	 		\langle a \rangle_{\epsilon}=\frac{d s_{0,\epsilon}}{d \beta}_{\beta=0}=-\frac{\partial F_{\epsilon}/\partial\beta}{\partial F_{\epsilon}/\partial s_{0,\epsilon}}\|_{\beta=0}\,,
 		\end{eqnarray}
 		where $\langle a\rangle_{\epsilon}$ is the perturbed observable average and $\langle a\rangle_{\epsilon=0}\equiv\langle a\rangle$ is the original one. 
 		
		The idea of perturbing chaotic systems may seem unreliable in regards with the presence of the dense set of periodic windows in the parameter space, but POT provides us with an intuitive theoretical framework to evaluate various perturbations. From Eqs.~\ref{Formula:perturbed dynamical zeta function} and \ref{Formula:average computing}, it is clear that the continuous deformation of an individual cycle $p$ changes $F_{\epsilon}$ and its derivatives smoothly, and thus the observable average $\langle a \rangle_{\epsilon}$ is an analytic function of $\epsilon$ for a finite truncation if all the involved cycles continue existing. Even weak perturbations may be classified into two basic types: the ones that maintain the symbolic dynamics and those that result in birth or death of cycles. Both types could lead to displacement and deformation of the UPOs while the latter ones further lead to creation or annihilation of UPOs and loss of analyticity of Eq.(\ref{Formula:perturbed dynamical zeta function}) as an infinite product. Nevertheless, if the pruning rule is known as $\epsilon$ varies, the analyticity could still be utilized for each cycle that continues to exist throughout, which still results in a good approximation as we will see in the following . Hence, with different $\epsilon$'s, the amount of calculation will be greatly reduced if we have the qualitative knowledge of the influence of the perturbation on the existence of cycles. From this standpoint, cycle expansions give us inspiration to quantify the perturbation on chaos.
		\subsection{\label{subsec:Perturbations in the Complex Plane}Perturbations in the Complex Plane}
		As discussed in Sect.~\ref{subsec:Perturbed Model}, we evaluate the observable average with a given perturbation $\epsilon g(x)$. To accomodate the continuous change of $\epsilon$, a natural and efficient approach is to perform a series expansion. For a prefixed $g(x)$, $\langle a\rangle_{\epsilon}$ can be viewed as a function of $\epsilon$. With a proper selection of cycles, the average at a ``target'' $\hat{\epsilon}$ based on the values at $\epsilon=\epsilon_0$ could be written as
			\begin{eqnarray}{\label{Formula:Taylor expansions on <a>}}
				\langle \hat{a}\rangle_{\hat{\epsilon}}=\langle a\rangle_{\epsilon=\epsilon_0}+(\hat{\epsilon}-\epsilon_0)\frac{d\langle a\rangle_{\epsilon}}{d\epsilon}|_{\epsilon=\epsilon_0}+\frac{(\hat{\epsilon}-\epsilon_0)^2}{2!}\frac{d^2 \langle a\rangle_{\epsilon}}{d\epsilon^2}|_{\epsilon=\epsilon_0}+\frac{(\hat{\epsilon}-\epsilon_0)^3}{3!}\frac{d^3 \langle a\rangle_{\epsilon}}{d\epsilon^3}|_{\epsilon=\epsilon_0}+...\,,
			\end{eqnarray}
			where the accuracy of $\langle \hat{a}\rangle_{\hat{\epsilon}}$ depends on the order and accuracy of its derivatives we evaluate. It has to be emphasized that now we assume that $\epsilon$ changes in a direction that maintains or reduces cycles and the series expansion of $\langle \hat{a}\rangle_{\epsilon}$ is only performed with the cycles that continue to exist at $\hat{\epsilon}$, which requires extra effort to judge. The derivatives could be conveniently evaluated with parameter values slightly different from $\epsilon_0$ on the complex plane, to be explained below.
			
			For hyperbolic maps with complete binary symbolic dynamics, Eq.(\ref{Formula:perturbed dynamical zeta function}) as a $0$th-order approximation of Eq.(\ref{Formula:spectral determinant}) is an exponentially convergent infinite product over UPOs and the observable average $\langle a\rangle_{\epsilon}$ related to the leading eigenvalue $s_{0,\epsilon}$ can be viewed as an analytic function in the complex-$\epsilon$ plane. Thus, an effective approach is to evaluate the derivative of $\langle a\rangle_{\epsilon}$ through the Cauchy integral formula~\cite{henrici1993applied}
			\begin{eqnarray}{\label{Formula:Cauchy integral formula}}
			\frac{d^k \langle a\rangle_{\epsilon}}{d\epsilon^k}|_{\epsilon=\epsilon_0}=\frac{k!}{2\pi i}\oint_{|r|<|\hat{\epsilon}-\epsilon_0|}\frac{\langle a\rangle_{\epsilon}}{(\epsilon_0-\epsilon)^{k+1}}d\epsilon\,,
			\end{eqnarray}	
			where we evaluate all the $\langle a\rangle_{\epsilon}$ along the circular integration path which encircles $\epsilon_0$ in the anti-clockwise direction on the complex plane and $|r|=|\epsilon-\epsilon_0|$ is a chosen integration radius which is usually smaller than $|\hat{\epsilon}-\epsilon_0|$. Of course, on the $\epsilon$-complex plane, the dynamics $f_{\epsilon}$ and the periodic points of the UPOs are all extended from the original ones (detailed in App.~\ref{appendix:extension to complex domains}). It should be noted that if a UPO is pruned at the chosen $\hat{\epsilon}$, it will not be included in the computation of $\langle a\rangle_{\epsilon}$ in Eq.(\ref{Formula:Cauchy integral formula}). Therefore, rigorously, the expansion Eq.(\ref{Formula:Taylor expansions on <a>}) holds only when there is no creation or annihilation of cycles. Otherwise, it has to be taken into account as just noted. 
		\subsection{\label{subsec:Perturbation While Pruning}Perturbation While Pruning}
			In certain cases, some symbol sequences have to be pruned~({\em e.g.}, Fig.~\ref{graphic:height case}.(a)) that correspond to non-existing orbits. To maintain the consistency and analyticity of the formulas and evaluate the derivatives of $\langle a\rangle_{\epsilon}$ reliably, before our computation, the prime cycles need to be judged to be admissible (discussed in Sect.~\ref{subsec:SD}) and all the inadmissible ones at a chosen $\hat{\epsilon}$ have to be eliminated. In one dimension, this could be done by the kneading theory while in two dimensions, the pruning front is a useful tool~\cite{cvitanovic1988topological}. Nevertheless at different points along the integration path, cycle expansions only involve those prime cycles that continue to exist up to the ``target'' perturbation. This judgement step does increase our computational effort, but it is clearly much more advantageous than the possible huge amount of computation involved in a direct application of the cycle expansions in the presence of continuously varying parameters.
			
			Next, we introduce the concept of covering map. A covering map covers the whole phase space, which admits the full symbolic dynamics. That is, all the symbolic prime cycles may be matched with admissible UPOs. Even if the map is covering at a particular $\epsilon_0$, on the complex-$\epsilon$ plane, it is possible that some cycles may get pruned along the integration path around $\epsilon=\epsilon_0$, which should be avoided. Very often, the problem could be fixed with a new choice of the perturbation center $\epsilon_0'$ or a new integration path. One good thing about the current scheme is that once the covering map with the complex parameters are found, it could be utilized throughout the whole computation. If some cycles are pruned at $\epsilon=\hat{\epsilon}$, we just do not include them in the calculation of $\langle a\rangle_{\epsilon}$ when evaluating the derivatives with Eq.(\ref{Formula:Cauchy integral formula}). Thus, if the pruning rule could be figured out when parameters are varying, the covering map could be conveniently used to compute dynamical averages of any smooth observables. Of course, if the symbolic dynamics remains unchanged during the parameter variation, the average $\langle a\rangle_{\epsilon}$ becomes truly analytic and Eq.(\ref{Formula:Taylor expansions on <a>}) holds for all parameters on the variation path. All the involved derivatives need just evaluating once. 
	\section{\label{sec:Examples}Examples}
		Based on cycle expansions, we demonstrate the perturbation scheme when varying a parameter in chaotic systems. In view of the two types of perturbation proposed in Sect.~\ref{subsec:Perturbed Model}, we apply the scheme to compute the observable averages of the following perturbed models to verify its effectiveness. Before doing that, some details in numerical computation need to be noted. 
		\subsection{\label{subsec:Some notes on numerical computation}Some Notes on Numerical Computation}
			To integrate along the path in Eq.(\ref{Formula:Cauchy integral formula}), it is necessary to introduce a feasible discrete scheme. In the following computation, the circular integration path is sampled regularly and the $m$ lattice points are named sequentially as $\{\epsilon_{r,i}\},i=1,2,3,...m$. Further, $d\epsilon$ is replaced by $\Delta\epsilon_i=\epsilon_{r,i+1}-\epsilon_{r,i}$ when $i=1,2,3,...m-1$ and $\Delta\epsilon_{r,m}=\epsilon_{r,1}-\epsilon_{r,m}$. Then we approximate Eq.(\ref{Formula:Cauchy integral formula}) with a summation 
			\begin{eqnarray}{\label{Formula:Cauchy summation formula}}
			\frac{d^k \langle a\rangle_{\epsilon}}{d\epsilon^k}_{\epsilon=\epsilon_0}=\frac{n!}{2\pi i}\sum_{i=1}^{m}\alpha_i\frac{\langle a\rangle_{\epsilon_{mid,i}}}{\epsilon_{mid,i}^{k+1}}\Delta\epsilon_i=\frac{n!}{2\pi i}\sum_{i=1}^{m}\frac{\alpha_i\langle a\rangle_{\epsilon_{mid,i}}\Delta\epsilon_i}{\epsilon_{mid,i}^{k+1}}\,,
			\end{eqnarray}
			where the point $\epsilon_{mid,i}=\frac{\epsilon_{r,i}+\epsilon_{r,i+1}}{2}$ is the midpoint of the $i-$th edge and $\alpha_i=1+\frac{(-1)^i}{3}$ are set according to the Simpson's rule. All the $\langle a\rangle_{\epsilon_{r,\cdot}}$'s are obtained with Eq.(\ref{Formula:perturbed dynamical zeta function}) and the corresponding complex UPOs. The radius $|r|=|\epsilon-\epsilon_0|$ of our chosen integration path should not be too small which could lead to large errors in the evaluation of high order derivatives, but usually needs to be small enough to ensure that all prime cycles exist along the integration path. As shown in Fig.~\ref{graphic:shift case}.(d), the larger $m$ is, the more accurate this approximation is. The averages $\langle a\rangle_{\hat{\epsilon}}$ obtained by a direct application of Eq.(\ref{Formula:perturbed dynamical zeta function}) are ``target values'' in comparison with the predicted averages $\langle \hat{a}\rangle_{\hat{\epsilon}}$ to assess the accuracy of our scheme. Due to the limited precision, our computation yields complex values with small imaginary parts which are also an indicator of the calculation accuracy. In addition, the values obtained through the Monte Carlo method~\cite{metropolis1949monte} are used as a benchmark to compare the ``target values''. In some cases, the direct calculation tends to converge slowly and is not as accurate as the Monte Carlo one, but we still use the ``target values'' for comparison and to evaluate the results of the new scheme. 
			
			In the following examples, we will state each chosen circular integration path and the regular and predict the observable averages with the Taylor expansion Eq.(\ref{Formula:Taylor expansions on <a>}), where the series are kept up to the 6th-order for good accuracy. Both computational accuracy and efficiency are considered for setting the truncation length $L_{max}$ for cycle expansions. Different $L_{max}$ are set in different examples to adapt to different convergence rates. A large $L_{max}$ is employed when the convergence is slow. With a good Markov partition of the phase space, the symbolic dynamics could be used to mark admissible cycles~\cite{hao1998applied,kitchens2012symbolic}. The multiple shooting method~\cite{cvitanovic2005chaos} is very effective in this case and thus used to search cycles. In addition, it is useful to note that in different examples, $\epsilon$ appears in different locations to indicate different types of perturbations while the current scheme applies to all different cases.
		\subsection{\label{subsec:Perturbations Maintaining Symbolic Dynamics}Perturbations Maintaining Symbolic Dynamics}
\begin{figure}[H]
				\centering
				\subfigure[]{\includegraphics[scale=0.5]{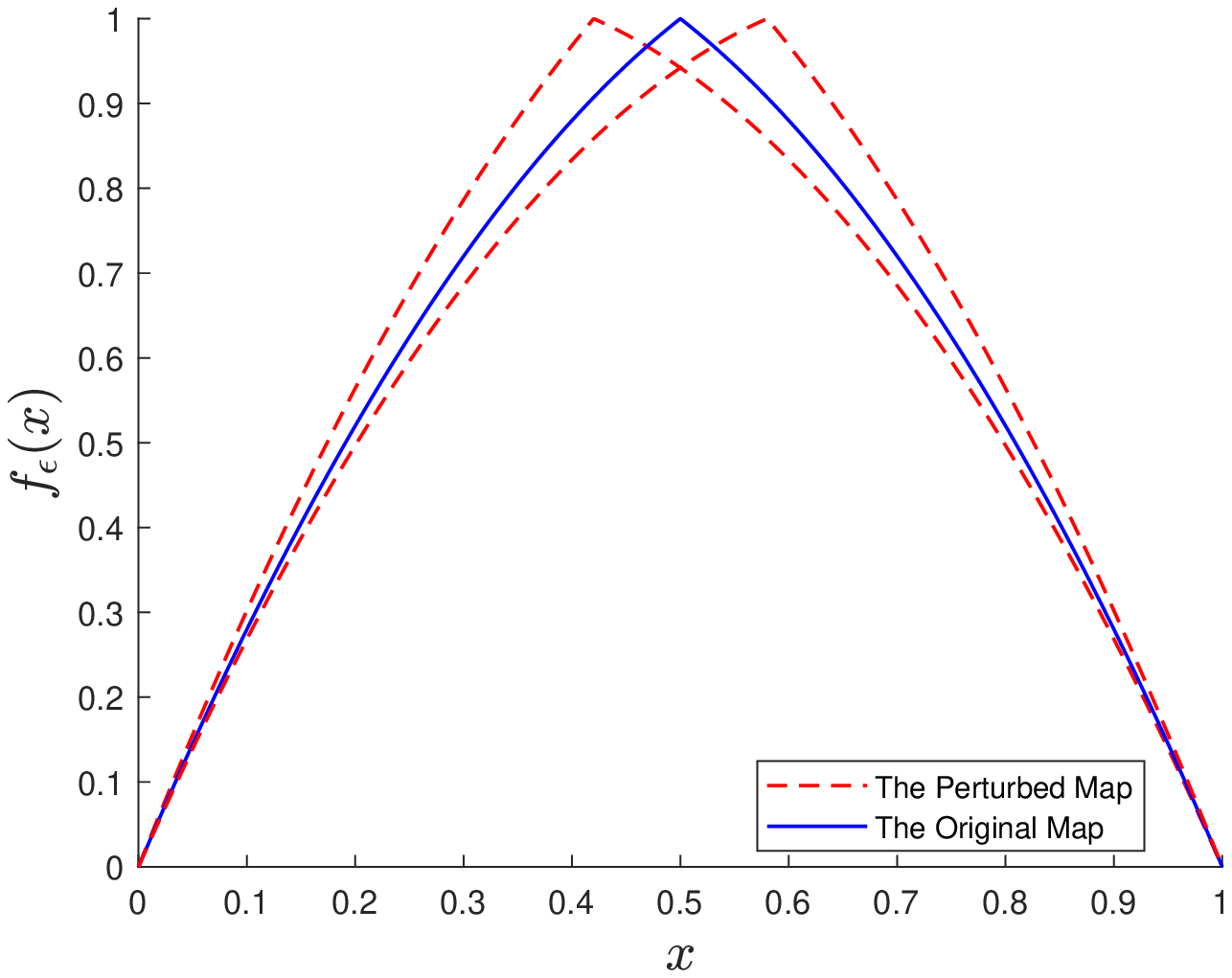}}
				\subfigure[]{\includegraphics[scale=0.5]{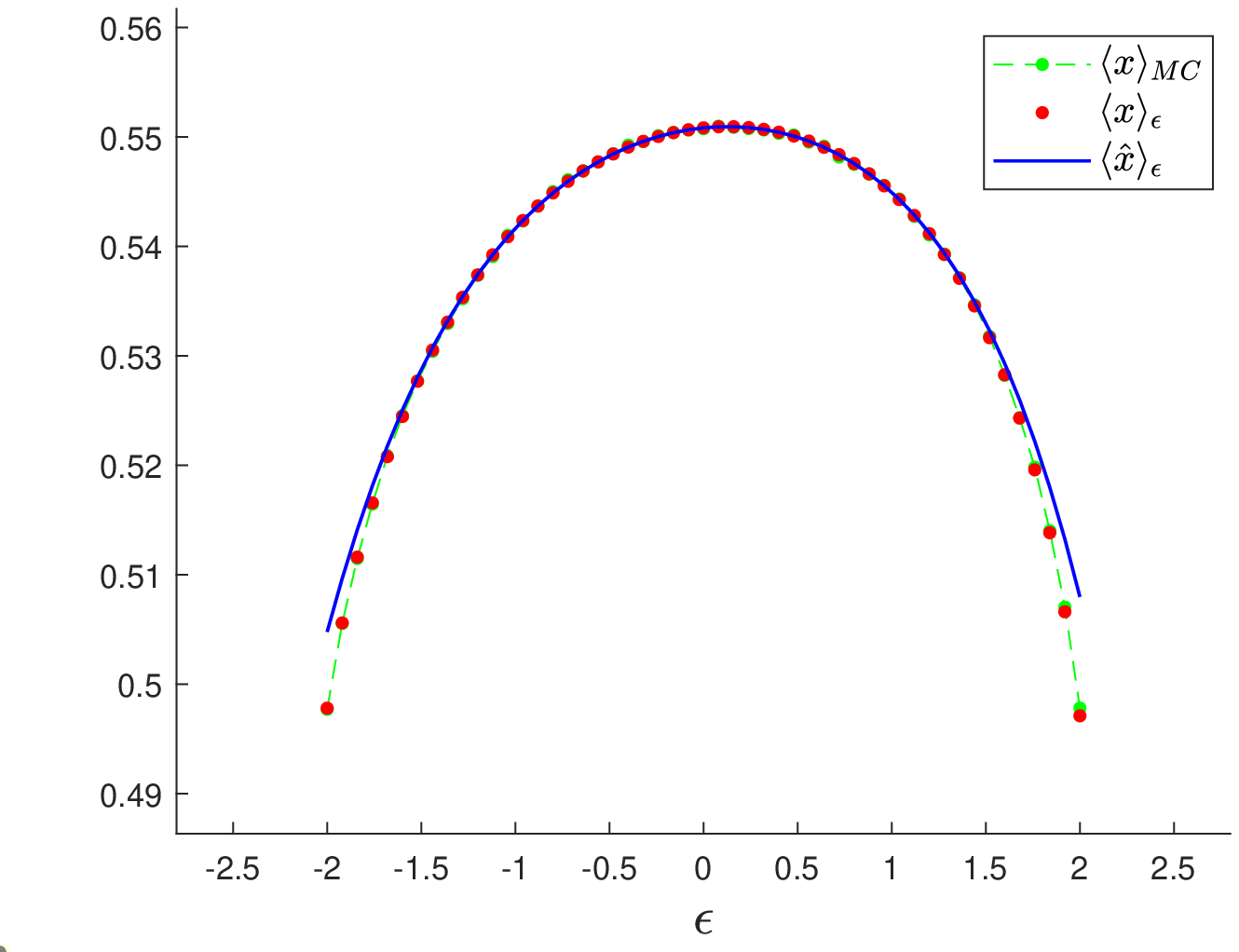}}
				\subfigure[]{\includegraphics[scale=0.5]{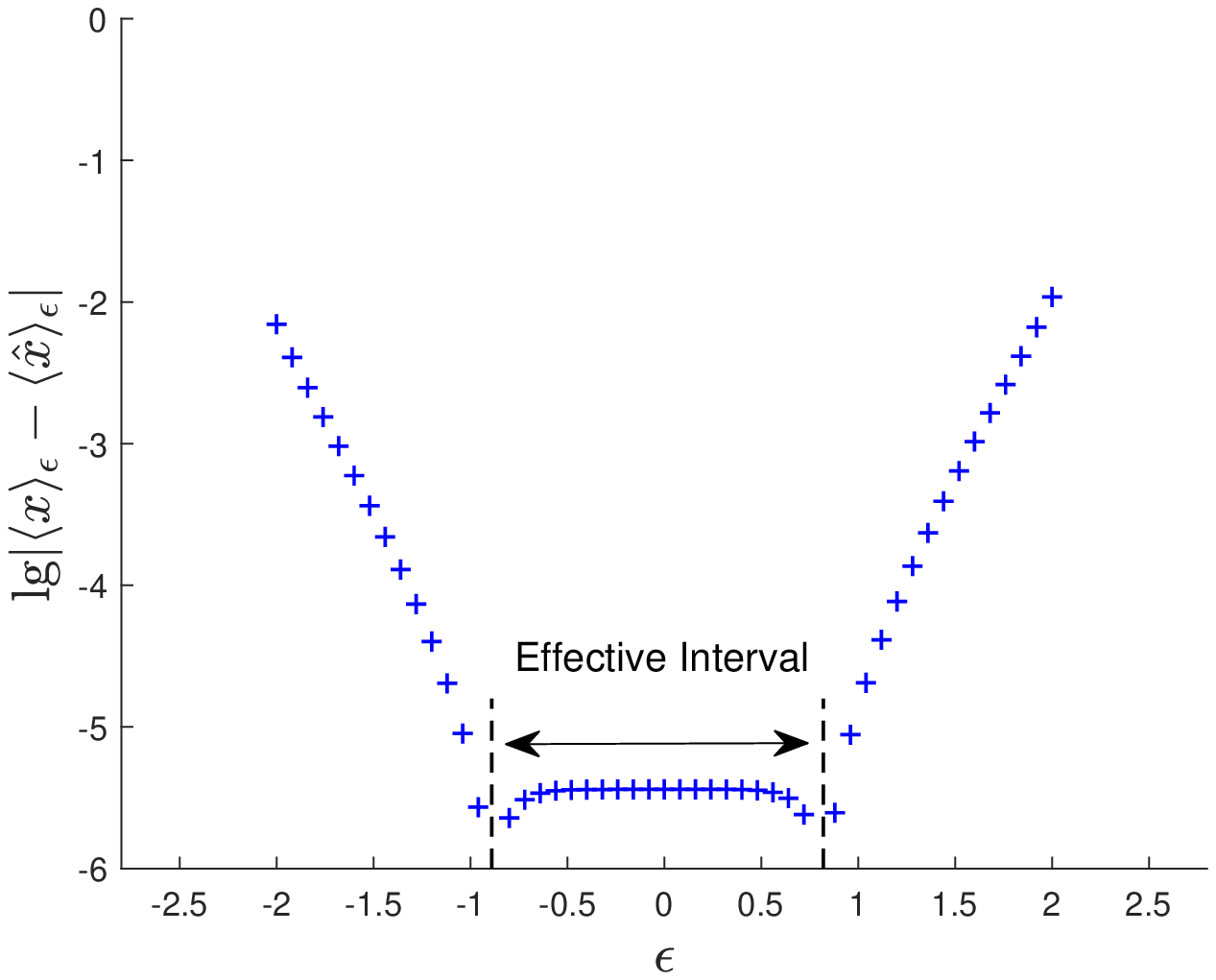}}
				\subfigure[]{\includegraphics[scale=0.5]{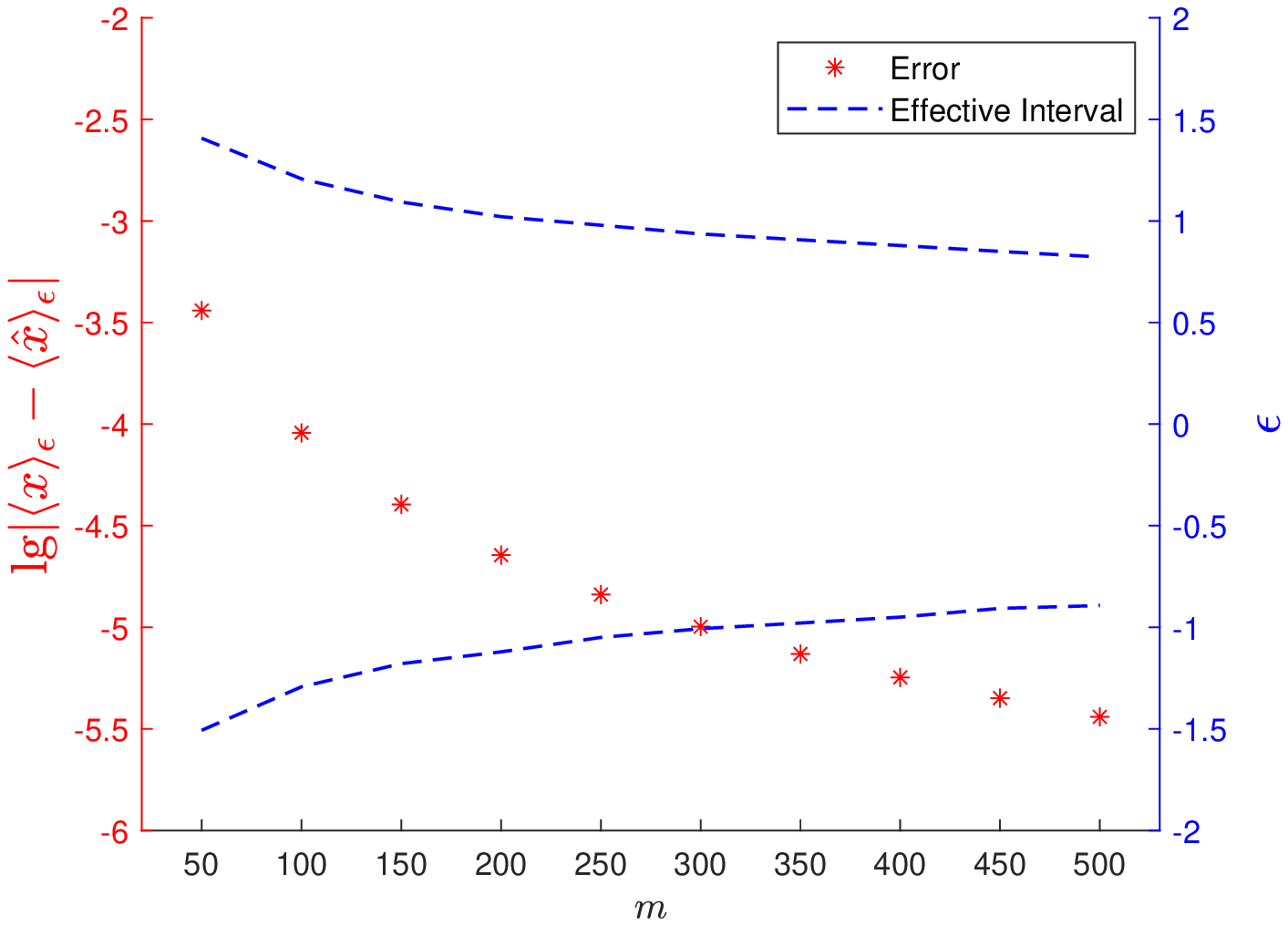}}
				\caption{A perturbed tent-like map which maintains the complete symbolic dynamics. (a) The original map (blue solid line, $\epsilon=0$) and the perturbed ones (red dashed lines, $\epsilon=-0.8$ and $0.8$). (b) the predicted values $\langle \hat{x}\rangle_{\epsilon}$ (blue line) are in good agreement with the directly calculated values $\langle x\rangle_{\epsilon}$ (red dots) and the Monte Carlo results $\langle x\rangle_{MC}$ (green dashed line) for given $\epsilon$-perturbations. (c) the errors between the predicted values $\langle \hat{x}\rangle_{\epsilon}$ and the directly calculated values $\langle x\rangle_{\epsilon}$ could reach $10^{-5.4}$ in the effective interval (roughly the interval between the black dashed lines). (d) the approximation Eq.(\ref{Formula:Cauchy summation formula}) gains accuracy (red star) as the number $m$ of sampling points increases while the applicability range (the blue dashed lines indicate the change of the interval endpoints) of our method narrows.}
				\label{graphic:shift case}
			\end{figure}

			If a perturbation slightly deforms the UPOs but maintains the symbolic dynamics, we may directly apply the perturbation expansion Eq.(\ref{Formula:Taylor expansions on <a>}) and (\ref{Formula:Cauchy integral formula}) for different values of $\hat{\epsilon}$. Furthermore, if the distribution of $\hat{\epsilon}$ is known in this case, another average with respect to $\hat{\epsilon}$ could be done easily. The famous tent map is not a good choice to be used as an demonstration here, because the uniform natural measure and the observable averages do not depend on the position of the critical point. Instead, we use a slightly altered tent-like model to validate our method in a simple case 
			\begin{eqnarray}{\label{Formula:the shift map}}
			f_{\epsilon}(x)=\begin{cases}
			-2x^2+\frac{\epsilon^2+10\epsilon+75}{25+5\epsilon}x\,&x\in[0,x_{c,\epsilon}]\\
			-2x^2-\frac{\epsilon^2+10\epsilon-25}{25-5\epsilon}x+\frac{\epsilon^2+25}{25-5\epsilon},&x\in(x_{c,\epsilon},1] 
			\end{cases}\,,
			\end{eqnarray}
			where $\epsilon$ controls the degree of deformation and the critical point $x_{c,\epsilon}=\frac{5+\epsilon}{10}$ moves as $\epsilon$ varies while the function value is always $1$ at this point (the perturbed tent-like maps with $\epsilon=-0.8,0$ and $0.8$ are shown in Fig.~\ref{graphic:shift case}.(a)). We choose the perturbation center at $\epsilon_0=0$, the number of sampling points along the integration contour is $m=500$, with the integration radius $r=0.1$ and the truncation length $L_{max}=10$. According to Eqs.~\ref{Formula:Taylor expansions on <a>} and \ref{Formula:Cauchy summation formula}, we have
			\begin{eqnarray}{\label{Formula:Example of Taylor expansions}}
			\langle \hat{a}\rangle_{\hat{\epsilon}}=\langle a\rangle_{\epsilon=0}+\frac{\,\hat{\epsilon}}{2\pi i}\sum_{i=1}^{500}\frac{\alpha_i\langle a\rangle_{\epsilon_{r,i}}\Delta\epsilon_i}{\epsilon_{r,i}^{2}}+\frac{\,\hat{\epsilon}^2}{2\pi i}\sum_{i=1}^{500}\frac{\alpha_i\langle a\rangle_{\epsilon_{r,i}}\Delta\epsilon_i}{\epsilon_{r,i}^{3}}+\frac{\,\hat{\epsilon}^3}{2\pi i}\sum_{i=1}^{500}\frac{\alpha_i\langle a\rangle_{\epsilon_{r,i}}\Delta\epsilon_i}{\epsilon_{r,i}^{4}}+...\,,
			\end{eqnarray}
			where the ``target'' $\hat{\epsilon}$ actually refers to each point in the ``target interval'' $[-2,2]$ in this example and $\{\epsilon_{r,i}\}$ are uniformly distributed on the integration path. Fortunately, the prediction $\langle \hat{x}\rangle_{\epsilon}$ can match the ``target values'' and the Monte Carlo results very well in the effective interval $\epsilon\in[-0.9,0.8]$ while the prediction is less accurate beyond this range. Actually, an increase in error outside the effective range is a reasonable phenomenon for perturbation approximation. For higher accuracy, Eq.(\ref{Formula:Taylor expansions on <a>}) needs to be extended to higher orders. The comparison among different methods is made in Fig.~\ref{graphic:shift case}.(b). The errors $\mathrm{log}|\langle \hat{x}\rangle_{\epsilon}-\langle x\rangle_{\epsilon}|$ are evaluated (Fig.~\ref{graphic:shift case}.(c)) to show the accuracy of our predictions which can reach $10^{-5.4}$ for reasonable perturbations. The improvement in accuracy shown at the two endpoints of the effective interval is due to the fact that the systematic error is accidentally compensated by the increase of the predicted values of the observables away from $\epsilon_0$. As we can see in Fig.~\ref{graphic:shift case}.(d), the computation gains accuracy as $m$ increases while the effective range becomes narrowing down. This example illustrates the effectiveness of our scheme under a weak perturbation which keeps the symbolic dynamics invariant, and the ability to adjust the computational parameter ({\em i.e.}, the number of sampling points $m$) to meet the accuracy requirements.
		\subsection{\label{subsec:Perturbation that Induces Pruning}Perturbation that Induces Pruning}
\begin{figure}[H]	
				\centering
				\subfigure[]{\includegraphics[scale=0.5]{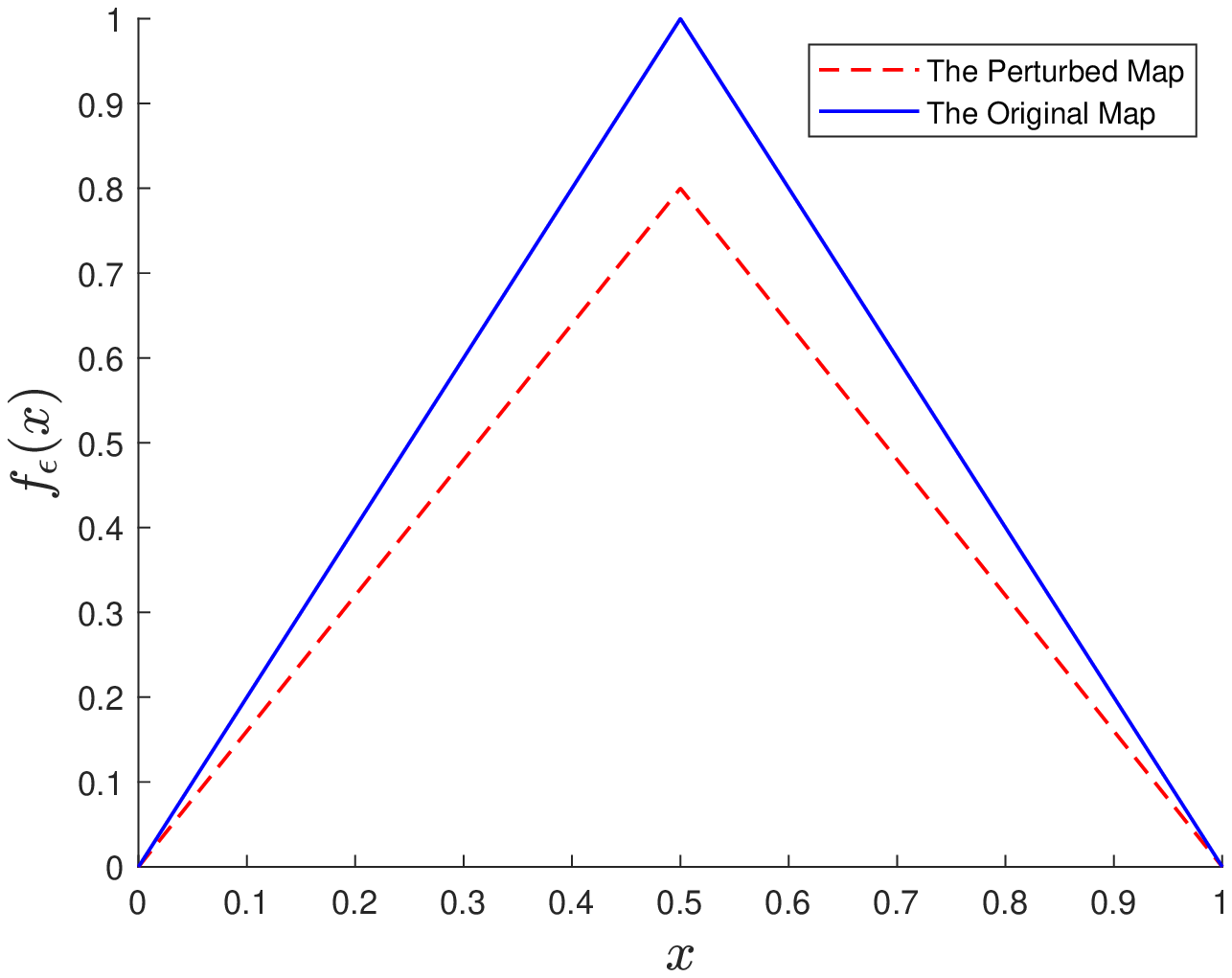}}
				\subfigure[]{\includegraphics[scale=0.5]{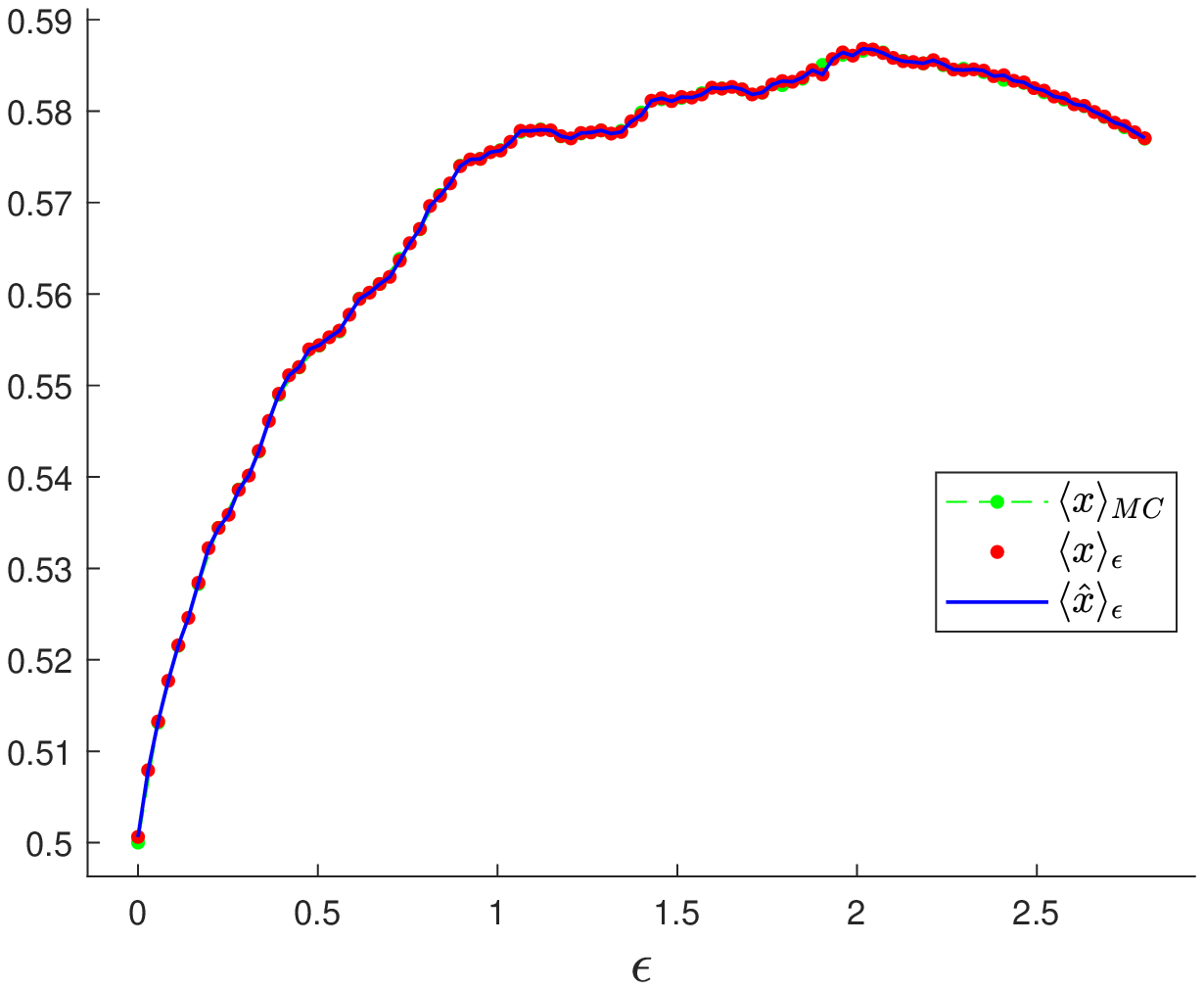}}
				\subfigure[]{\includegraphics[scale=0.5]{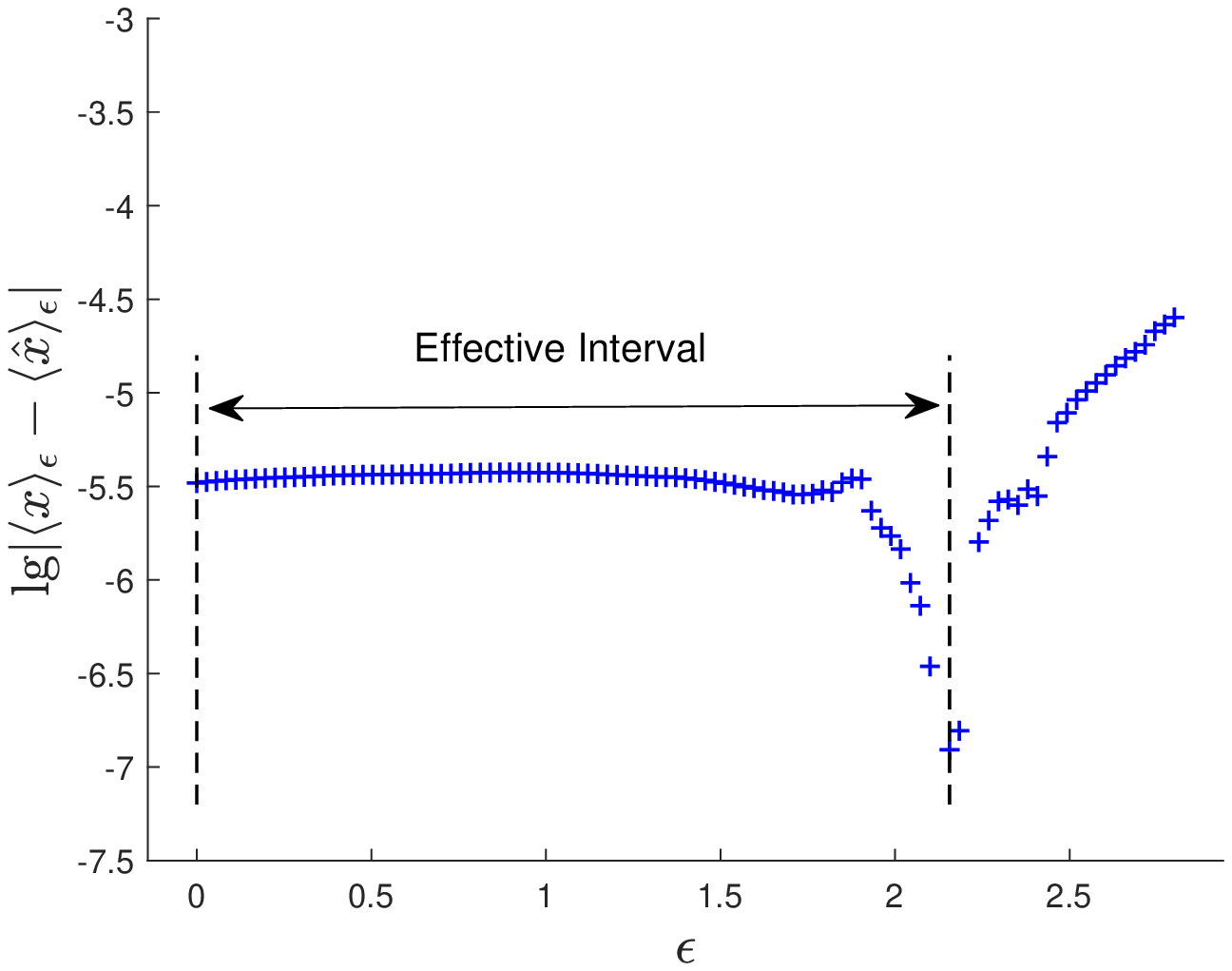}}
				\caption{The tent map with a perturbation that induces pruning. (a) The original map (blue solid line, $\epsilon=0$) and a perturbed one (red dashed line, $\epsilon=1$). (b) the predicted values $\langle \hat{x}\rangle_{\epsilon}$ (blue line) match the ``target values'' $\langle x\rangle_{\epsilon}$ (red dots) and the Monte Carlo results $\langle x\rangle_{MC}$ (green dotted line) with an expected and improvable accuracy in the effective interval. (c) the errors between the predicted values $\langle \hat{x}\rangle_{\epsilon}$ and the ``target values'' $\langle x\rangle_{\epsilon}$ which could reach $10^{-5.5}$ in the interval of applicability (roughly between the black dashed lines).}
				\label{graphic:height case}
			\end{figure}

			Usually, perturbed chaotic systems that maintain the symbolic dynamics are rare, and the annihilation or creation of UPOs invariably occurs. For instance, we consider a simple tent map the peak height of which changes as $\epsilon$ varies  
			\begin{eqnarray}{\label{Formula:the height map}}
				f_{\epsilon}(x)=\begin{cases}
				(2-0.2\epsilon)x\,&x\in[0,1/2]\\
				(2-0.2\epsilon)(1-x),&x\in(1/2,1] 
				\end{cases}\,,
			\end{eqnarray}
			where $\epsilon$ marks the strength of the perturbation. When $\epsilon<0$, the peak is beyond the phase space $[0,1]$ which leaves some trajectories quickly escaping but this situation has complete symbolic dynamics and thus could be treated in a way similar to the previous example. Here the real concern is the pruning case with $\epsilon>0$ and the perturbed tent maps with $\epsilon=0$ and $1$ are shown in Fig.~\ref{graphic:height case}(a). As discussed in Sect.~\ref{subsec:Perturbation While Pruning}, we need to evaluate in advance which UPOs should be pruned before predicting the observable averages at the chosen $\hat{\epsilon}$. Intuitively speaking, any UPO that visits the pruning interval $(f_{\epsilon}(1/2),1]$ is inadmissible. Within the set truncation length $L_{max}=12$, for instance, there are $747$ UPOs at $\epsilon=0$ while $504$ UPOs are pruned at $\epsilon=1$. In the practical computation, importantly, we must ensure that the coverage along the integral path includes the coverage at the chosen $\hat{\epsilon}$, so that all the prime cycles at $\hat{\epsilon}$ also exist on the path. Thus, we expand $\langle x\rangle_{\epsilon}$ at $\epsilon_0=-0.15$ and compute the derivatives along a new integration path $|\epsilon-\epsilon_0|=0.1$ with $m=500$, and the latter steps are unchanged. As shown in Figs.~\ref{graphic:height case}.(b) and (c), the observable averages $\langle \hat{x}\rangle_{\epsilon}$ in the interval $\epsilon\in[0,3]$ match the ``target values'' and the Monte Carlo results with an expected and further improvable accuracy.
		\subsection{\label{subsec:Perturbing 2-D Model}Perturbing 2-dimensional Model}
\begin{figure}[H]	
				\centering
				\subfigure[]{\includegraphics[scale=0.5]{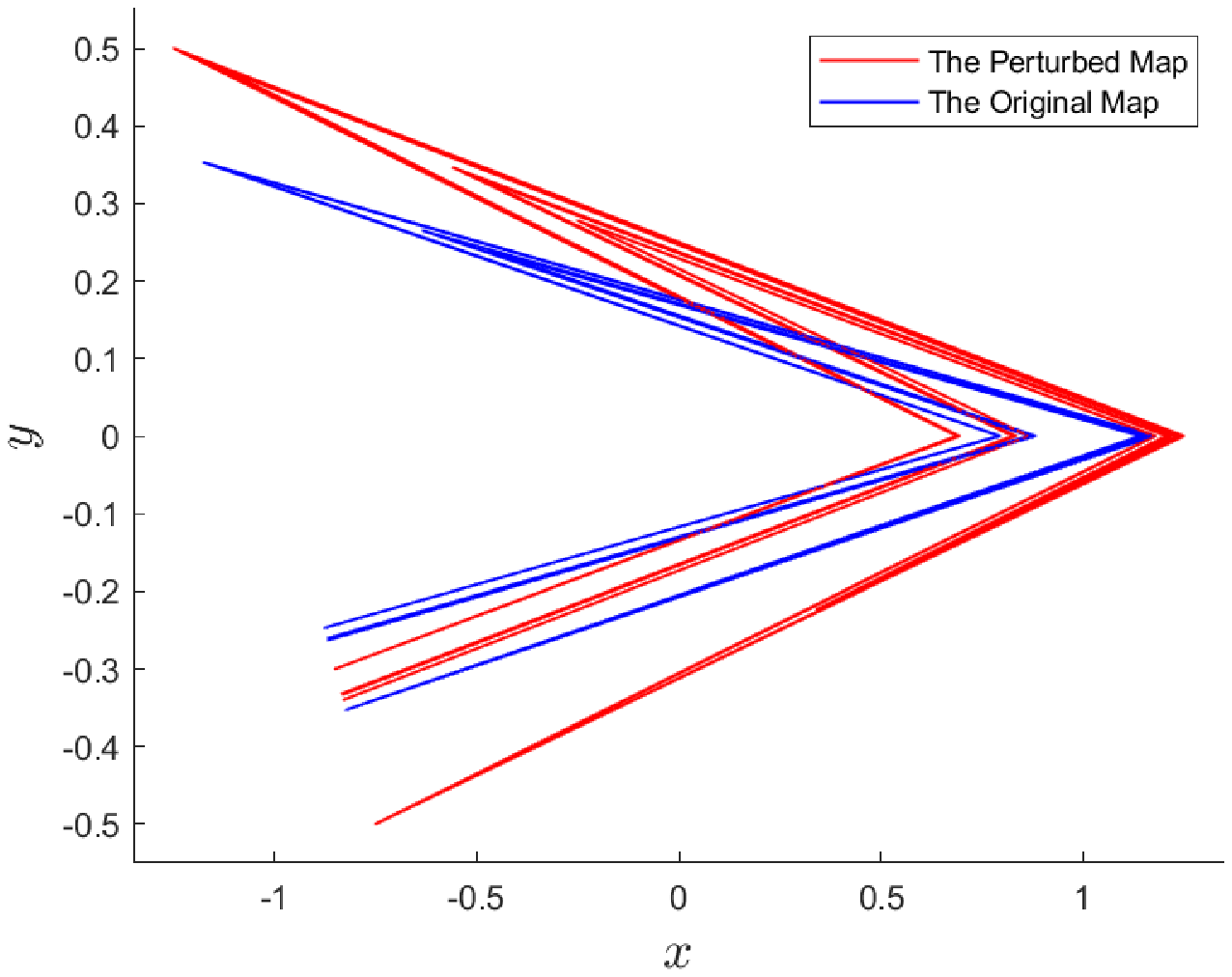}}
				\subfigure[]{\includegraphics[scale=0.5]{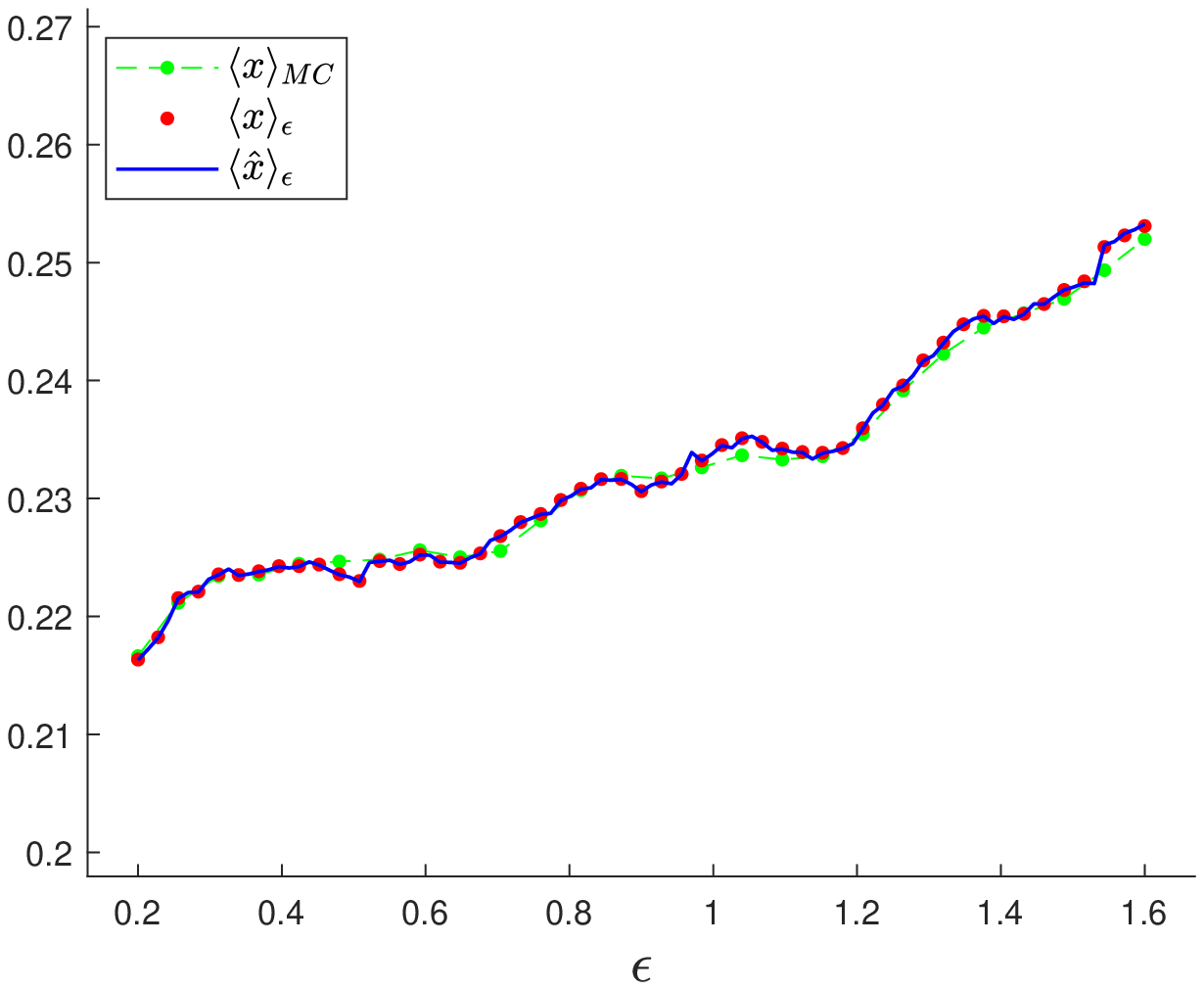}}
				\subfigure[]{\includegraphics[scale=0.5]{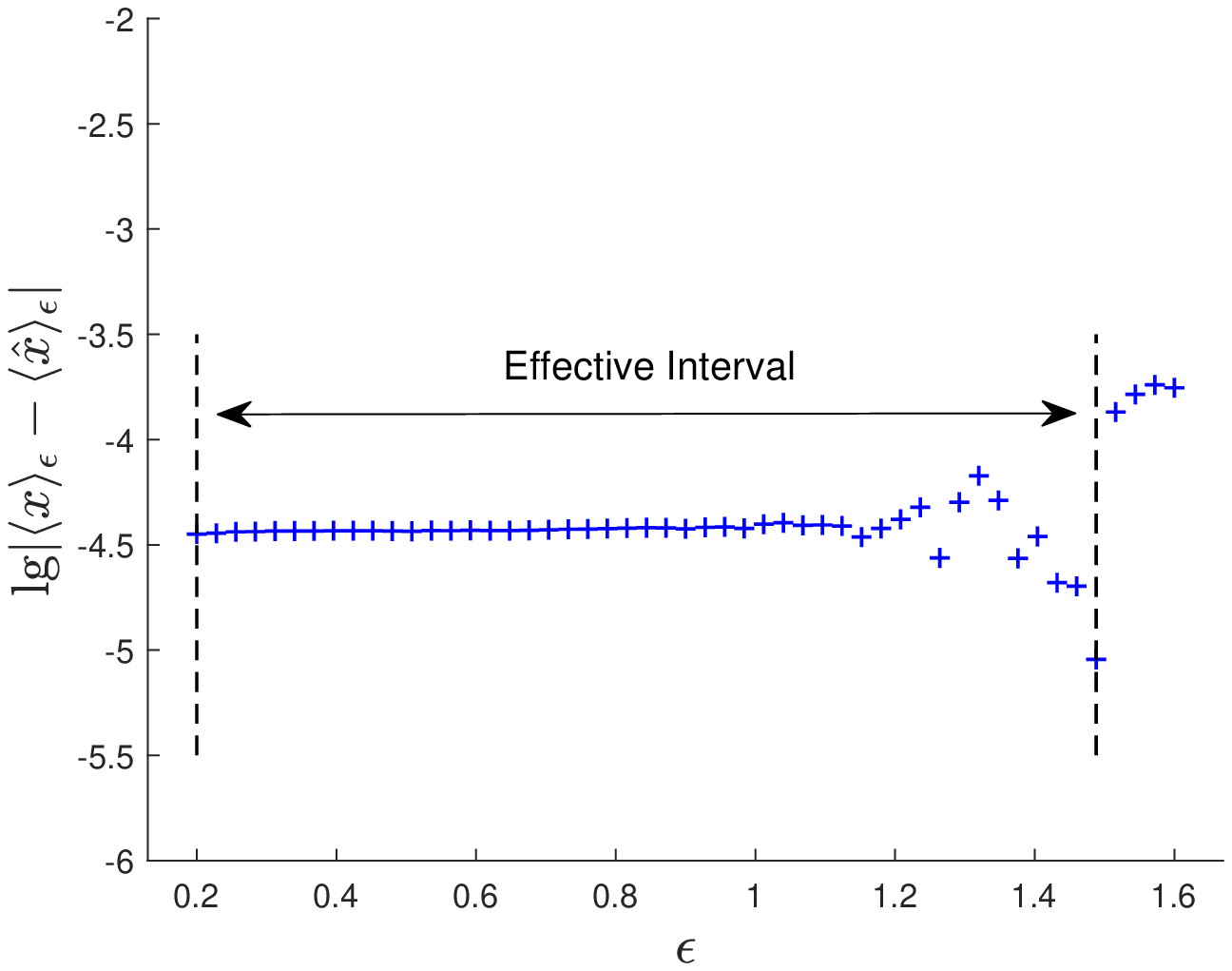}}
				\caption{Computation on the perturbed lozi map: (a) the attractor images of the lozi map with parameters $(1.8,0.4)$ (blue line) or $(1.85,0.3)$ (red line). (b) the predicted values $\langle \hat{x}\rangle_{\epsilon}$ (blue line) agree with the ``target values'' $\langle x\rangle_{\epsilon}$ (red dots) with expected accuracy while slight error exists between the target values and the Monte Carlo results (green dotted line). (c) the logarithmic errors between the predicted values $\langle \hat{x}\rangle_{\epsilon}$ and the ``target values'' $\langle x\rangle_{\epsilon}$ which prove the validity of the perturbation scheme in this 2-dimensional model.}
				\label{graphic:lozi case}
			\end{figure}

			The previous two examples demonstrate the effectiveness of our scheme in dealing with the two types of perturbations in one-dimensional maps. We further validate our scheme in a well-known two-dimensional model, the Lozi map~\cite{lozi1978attracteur}
			\begin{eqnarray}{\label{Formula:the lozi map}}
				(x,y)\mapsto f^{(a,b)}(x,y)=(1-a|x|+by,x)\,,
			\end{eqnarray}
			where $a$ and $b$ are adjustable parameters controlling the folding and stretching in the phase space. By varying the parameters, the structure of the invariant manifolds keeps changing and so does the strange attractor. A partition of the plane by the y-axis determines the UPOs uniquely through the binary symbolic sequences. The parameters $a,b$ on the crisis line $a=2-b/2$~\cite{tel1983invariant} are the largest values for which a strange attractor exists~\cite{cvitanovic1988topological} and there is a heteroclinic tangency at the intersection of the unstable manifolds of the ``$1$'' and ``$0$'' fixed point. As $b$ increases, some UPOs covering the phase space are gradually pruned and no new UPOs appear, so that our perturbation scheme can be applied. Here we set up a perturbed model  
			\begin{eqnarray}{\label{Formula:the perturbed lozi map}}
				(x,y)\mapsto f_{\epsilon}(x,y)=(1-\epsilon)f^{(a_1,b_1)}(x,y)+\epsilon f^{(a_2,b_2)}(x,y)\,,
			\end{eqnarray}
			where $(a_1,b_1,a_2,b_2)$ is set to $(1.85,0.3,1.8,0.4)$ defining a perturbation direction that we select along the crisis line. Eq.(\ref{Formula:the perturbed lozi map}) allows us to control the values of both $a$ and $b$ in Eq.(\ref{Formula:the lozi map}) with a single parameter $\epsilon$. The attractors corresponding to $\epsilon=0$ and $1$ are plotted in Fig.~\ref{graphic:lozi case}.(a). As $\epsilon$ increases, $b$ increases along the crisis line and $a$ decreases accordingly. In this computation, we expand $\langle x\rangle_{\epsilon}$ at $\epsilon_0=0.1$ and compute the derivatives along the integration path $|\epsilon-\epsilon_0|=0.1$ with an approximation $m=100$, and the cycle expansion is truncated at $L_{max}=17$. It is worth clarifying that the perturbation center and the integration path are not the only choice here. We just need to ensure that all the prime cycles at the target $\hat{\epsilon}$ exist on the integration path. As before, for any selected target $\hat{\epsilon}$, we should determine which prime cycles need to be pruned in advance as described in Sect.~\ref{subsec:SD}. The $\epsilon$-values are sampled from the interval $[0.2,1.6]$ and the results are plotted in Figs.~\ref{graphic:lozi case}.
			
			As shown in Figs.~\ref{graphic:lozi case}(b) and (c), the predicted values $\langle \hat{x}\rangle_{\epsilon}$ and the directly calculated ones $\langle x\rangle_{\epsilon}$ match well as expected. However, they do not agree very well with the Monte Carlo results. The discrepancy originates from the slow convergence of cycle expansion at some parameter values. Of course $\langle x\rangle_{MC}$ itself may not be so accurate. An increase of the truncation length may reduce the discrepancy. Actually, accelerating convergence is not the focus of this paper while the good agreement between the predictions and the ``target values'' already tells the validity of our perturbation scheme with cycle expansion in 2-dimensional models.
			
	\section{\label{sec:Summary}Summary}
		The main body of work in this paper is to verify our proposed perturbation calculation for chaotic systems. While the evolution of a single trajectory of a chaotic system is difficult to track, the POT states that the global behaviour of the system can be computed with UPOs which densely cover the phase space. In view of the smooth change of the UPOs before bifurcation, the dynamical zeta function (Eq.(\ref{Formula:perturbed dynamical zeta function})) varies analytically in a finite approximation based on these cycles so that the observable averages are amenable to simple Taylor expansion if the parameters do not change too much and the system remains chaotic.
		
		We propose a feasible scheme combining cycle expansions, analyticity and some necessary approximations to quantify the impact of perturbations on the statistical behaviour of chaotic systems. The scheme is detailed in the presence or absence of pruning upon parameter changes. Its effectiveness is demonstrated with several 1- or 2-dimensional models in Sect.~\ref{sec:Examples}. Of course, there are limitations in the computation, for example, the accuracy of our prediction depends on the convergence rate of the cycle expansion, etc. In fact, accelerating convergence is not the focus of this paper and a few acceleration schemes have been proposed in the literature~\cite{dettmann1997stability,artuso1990recycling2,cao2022wielding,gao2012accelerating}, but further discussion on how to integrate them into our scheme needs to be investigated. Furthermore, the admissibility criterion of UPOs in more complex systems also requires more discussion.

        During parameter changes, chaos attractors may lose stability to periodic motions and the chaotic trajectory turns transient. The current scheme is still valid but the obtained result is an average over the transient chaotic set. It is not the dynamical average obtained in a long-term simulation of the system and supported on the stable periodic orbit. Another complication is associated with the convergence rates of the Taylor and the cycle expansion. Although, in numerical computation, the valid range of perturbation parameter seems quite broad but we do not have a quantitative estimation of what it should be. On the other hand, even if chaotic motion is maintained some cycles may be on the edge of losing hyperbolicity, requiring more cycles to achieve high accuracy~\cite{artuso1990recycling1}. The good news is that if the system is nearly uniformly hyperbolic, this property will remain in a small perturbation of parameters and the current algorithm should work well. 
        		
		It is appropriate to say that the study of cycle expansions in perturbed chaotic systems is just started and far from complete. Therefore, we hope that this paper will give a taste of a new approach and provide a new tool to cope with perturbations in chaotic systems. Of course, exploring and promoting our scheme both in theory and in applications for higher-dimensional or even real systems requires further discussion and penetrating reflection.
	\begin{acknowledgements}
		This work was supported by the National Natural Science Foundation of China under Grants No.11775035, by BUPT Excellent Ph.D. Students Foundation, and also by the Key Program of National Natural Science Foundation of China (No. 92067202).
	\end{acknowledgements}
	\section*{Conflict of Interest}
		The authors declare that they have no conflict of interest.
	\section*{Data Availability Statement}
		All data, models, generated or used during the study appear in the submitted article, code generated during the study are available from the corresponding author by request.	
	
	\appendixpage
	\appendix
	\subsection{\label{appendix:SD}Pruning Algorithms in Sect.~\ref{subsec:SD}}	
		In 1-dimensional maps, the spatial ordering of a binary symbolic future itinerary $S^+=.s_1s_2s_3s_4...$ where $s_i \in \{0,1\}$ is converted to a binary number $\gamma(S^+)$, called future topological coordinate, by the converting algorithm~\cite{cvitanovic2005chaos}
		\begin{eqnarray}{\label{Formula:the converting algorithm}}
		\gamma(S^+)=\sum_{n=1}^{\infty}\frac{c_n}{2^n}\,,
		\end{eqnarray}
		where $c_{n+1}=s_{n+1}+(-1)^{s_{n+1}}c_n$ and $c_1=s_1$. The itinerary of the critical point $x_c$, the kneading sequence, is denoted as $S^+(x_c)$ which represents the upper bound of spatial order in the phase space, that is, any prime cycles whose spatial order exceeds $S^+(x_c)$ is inadmissible. Thus, an applicable admissibility criterion can be expressed as: all the realized prime cycles must satisfy the discriminant condition
		\begin{eqnarray}{\label{Formula:the discriminant condition}}
		\hat{\gamma}(p)\leq\gamma(S^+(x_c))\,,
		\end{eqnarray}
		where $\hat{\gamma}(p)$ is the maximal topological coordinate of prime cycle $p$, {\em e.g.}, $\hat{\gamma}(\overline{011})=max\{\gamma(011011011...),\gamma(101101101...),\\\gamma(110110110...)\}$. When promoted to 2-dimensional cases, the algorithm will also take into account the past itinerary. The spatial ordering of a binary symbolic past itinerary $S^-=...s_{-3}s_{-2}s_{-1}s_0.$ is converted to a binary number, the past topological coordinate~\cite{cvitanovic2005chaos}, as
		\begin{eqnarray}{\label{Formula:the delta converting algorithm}}
		\delta(S^-)=\sum_{n=1}^{\infty}\frac{d_{1-n}}{2^n}\,,
		\end{eqnarray}
		where $d_{n-1}=1-s_{n}+(-1)^{s_{n}+1}d_n$ and $d_0=s_0$. Thus, we can construct a symbol square $[\delta,\gamma]$ in which the admissible and the forbidden motions are separated by a ‘pruning front’ in the two-dimensional phase space, which is usually fractal and consists of the set of all primary turning points. All the realized prime cycles must be located in the admissible zones. Certainly, in physical or numerical experiments, only finite precision can be achieved and it is reasonable to choose an n-bit precision approximation (subshift of finite type)~\cite{boccaletti2000control}.
		
		In some cases, we are able to locate all the short admissible UPOs even without knowing the pruning rule as long as a symbolic partition of the phase space is achieved. We simply try all the possible sequences which will provide different initial guesses to cycle searching algorithms such as the multiple shooting method in Sect.~\ref{subsec:Some notes on numerical computation}. Certainly, many symbol sequences do not match any admissible orbits because we have not used the pruning rule. However, this sort of search will cover all the possible cases and will not miss any existing cycle. 
		
	\subsection{\label{appendix:extension to complex domains}Details of the extension of dynamics to complex domains}	
	 If Eq.(\ref{Formula:perturbed dynamical zeta function}) is analytic in $\epsilon$, it can be extended to the complex $\epsilon$-domain. Thus, the observable average $\langle a\rangle_{\epsilon}$ related to the leading eigenvalue $s_{0,\epsilon}$ can be viewed as an analytic function in the complex $\epsilon$-plane, which is the core of the perturbation scheme and used to compute coefficients of the Taylor expansion. Correspondingly, the dynamics of the system $f_{\epsilon}$ and the periodic points should all be extended to the complex domain according to the following rules:
	 \begin{itemize}
	 	\item The formula of the dynamics remains unchanged, except that the critical points on the real axis become critical lines perpendicular to the real axis in the complex domain.
	 	\item The periodic points of all the UPOs become complex and follow the dynamics of the system strictly. We search for UPOs in the complex domain similarly as in the real domain, but the admissibility is checked by the real part of the coordinates.
	 	\item To ensure the consistency and analyticity, Eq.(\ref{Formula:perturbed dynamical zeta function}) should be modified slightly in the complex plane. The denominator of $t_p$ denotes the stability of each UPO and should change analytically with $\epsilon$ to preserve the weight of each UPO in cycle expansion, {\em e.g.}, for the tent map with binary symbolic dynamics, $|\Lambda_{p,\epsilon}|$ in Eq.(\ref{Formula:perturbed dynamical zeta function}) should be modified to $(-1)^{s_1+s_2+...+s_{n_p}}\Lambda_{p,\epsilon}$ ($s_i \in \{0,1\}$) and the sign of $t_{p,\epsilon}$ corresponds to the topological property of cycle $p$.
	 \end{itemize}
	\bibliography{bibfile}
\end{document}